# Resonant dielectric amplification of superconductivity at the FeSe:$SrTiO_3$ interface

Krzysztof Kempa[1*] and Michael J. Naughton[1], & Hanno H. Weitering[2]

[1]Department of Physics, Boston College, Chestnut Hill, Massachusetts 02467, USA

[2]Department of Physics and Astronomy, University of Tennessee, Knoxville TN 37996, USA

*Corresponding Author: kempa@bc.edu

**Abstract**

Epitaxial monolayers of FeSe grown on $SrTiO_3$ (STO) and related polar substrates exhibit superconducting transition temperatures of 65–70 K, far above the $\sim$ 8 K value of bulk FeSe. Although coupling to interfacial Fuchs–Kliewer (FK) phonons is widely believed to contribute to this enhancement, the apparent universality of high-$T_c$ superconductivity across polar interfaces presents a striking puzzle. Here, we identify the lower-energy FK mode near 59 meV, associated with the bulk longitudinal optical resonance of STO, as a conduit through which the substrate dielectric response enhances pairing in FeSe. Within a phenomenological Migdal–Eliashberg framework, incorporating the measured dielectric response of STO selectively amplifies the pairing interaction near this resonance and raises the theoretical upper bound of $T_c$into the experimentally observed range. A similar enhancement is reproduced for FeSe on $BaTiO_3$, while related polar substrates such as $TiO_2$, $LaFeO_3$, and $EuTiO_3$ possess the same key electrodynamic ingredients. We refer to this effect as resonant anti-shielding (RAS). More broadly, RAS identifies the bulk electrodynamic response of the substrate as a pairing amplifier, opening a route toward designing superconducting heterostructures through control of their dielectric environment.

## Introduction

Interfaces between dissimilar materials can host emergent electronic states that are absent in their bulk constituents. A paradigmatic example is a single-unit-cell thick FeSe film grown on STO, where the superconducting transition temperature $T_c$ is dramatically enhanced relative to bulk FeSe.[1,2] This remarkable increase has motivated extensive efforts to uncover the microscopic origin of the enhanced Cooper pairing interaction.[3-9] A leading hypothesis attributes this enhancement, in part, to the coupling between FeSe electrons and interfacial Fuchs–Kliewer (FK) phonons of the substrate, particularly the ~92-96 meV surface optical mode.[3,6] Early support for this picture came from the observation of replica bands in angle-resolved photoemission spectroscopy, offset from the main bands by similar energies.[3,10-12] These features were interpreted as signatures of strong electron–phonon coupling and motivated forward-scattering models of superconductivity.[3,12] In such models, long-range electric fields generated by polar surface modes penetrate the FeSe layer, producing an interaction sharply peaked at small momentum transfer. In this forward-focused limit, even moderate coupling can significantly enhance $T_c$, and theoretical studies indicate that such phonons may either directly induce pairing or cooperate with intrinsic spin-fluctuation mechanisms without substantially altering the gap symmetry.[4,8,9]

The coupling is mediated by electric fields that decay away from the interface, making both their strength and spatial extent sensitive to structural details. Within the forward scattering model, $T_c$ is expected to depend on interface-specific parameters, including the spatial separation between the FeSe layer and the STO substrate, the (anisotropic) dielectric screening environment, bulk band bending and charge transfer, and the momentum width of the interaction.[3] These parameters vary across material systems and growth conditions. Nonetheless, FeSe monolayers grown on a range of polar substrates, including $SrTiO_3$(110),[13,11] $BaTiO_3$(001) (BTO),[14] $LaFeO_3$(001) (LFO),[15] and both rutile[16] and anatase $TiO_2$,[17] consistently produce $T_c$ values in the range 60-80 K. $EuTiO_3$ (ETO) provides an additional instructive example. Although its reported $T_c$ of about 30 K is based on transport measurements[18] and therefore cannot be directly compared with the spectroscopic pairing scales quoted for the other substrates, FeSe/ETO exhibits a substantial superconducting gap of $\sim$ 12meV,[18] compared with $\sim$ 15 meV for FeSe:STO,[2,19] as measured with angle-resolved photoemission. Since ETO and LFO are both antiferromagnetic, substrate magnetism does not appear to exert a dominant influence on interfacial superconducting pairing.

Electronically, these systems are remarkably similar. In all cases, the FeSe layer is electron doped by charge transfer from the substrate, leading to a Lifshitz transition in which hole pockets at the Brillouin zone center drop below the Fermi level and electron pockets at the M-point acquire additional charge.[2,3] This establishes a broadly similar electronic structure across these interfaces, upon which substrate-dependent interactions act. In fact, heavy electron doping appears to be a prerequisite for achieving high $T_c$,[5] consistent with studies of FeSe intercalation compounds.[20,21] On the phonon side, however, the situation is less uniform. Higher-energy FK modes, often emphasized in spectroscopic studies, occur at different energies across the various aforementioned substrates[22,15] and are sensitive to interface structure, doping, and temperature.[22,6,7] This variability reflects the fact that polar substrates support multiple optical phonon branches, giving rise to several FK surface modes with distinct microscopic character and coupling strengths (Table 1).

One phonon feature, however, is particularly robust across all superconducting FeSe/interface systems. As summarized in Table 1, the lower energy FK surface mode of the perovskite substrates STO,[22-25] LFO,[26] BTO,[22,27] and ETO[28,29] closely track the corresponding bulk longitudinal optical (LO) phonons, with the FK modes consistently appearing in the 57-59 meV range.[29-32] In $TiO_2$ polymorphs, the FK modes occur at somewhat different energies[33,34] but remain tied to the corresponding bulk longitudinal modes.[35-37] In all cases, this correspondence indicates that the lower FK mode reflects the bulk longitudinal dielectric resonance, *i.e.*, the frequency at which $\varepsilon(\omega) \rightarrow 0$. Interestingly, in an estimate of $T_c$ for FeSe:STO, it was suggested that this lower FK mode enhances superconductivity,[38] possibly into the experimental range $T_c$. By contrast, higher-energy FK modes are generally detuned from the corresponding LO resonances, and therefore less effective in exploiting dielectric enhancement, despite their prominence in spectroscopic observations. Thus, the dominant coupling channel may not be determined by the highest-frequency optical phonon, but by the longitudinal mode most closely associated with the bulk dielectric resonance (Table 1).

Efficient coupling further requires that longitudinal polarizations possess a finite projection onto the interface normal. In cubic STO, the relevant $F_{1u}$ modes possess polarization components along all crystallographic directions, ensuring a finite projection onto the interface normal even for orientations such as (110),[11,13] consistent with the observed robustness of superconductivity across different surface terminations (Table 1). Detailed density functional perturbation theory (DFPT) calculations for FeSe:STO interfaces further show that the presence of a $TiO_x$ interlayer significantly modifies the higher-energy longitudinal modes, while the lower FK/LO mode remains comparatively unchanged.[39] Consistent with this picture, the energy and damping of the lower FK mode remain largely unchanged upon deposition of an FeSe overlayer,[22] indicating that the resulting electron–phonon coupling at that energy reflects the intrinsic dielectric response of the substrate rather than interface-specific structure.

In this work, we argue that coupling to the bulk longitudinal dielectric response, via the lower energy FK mode, can provide a significant contribution to the enhancement of superconductivity, complementing the widely discussed role of surface FK phonons. By incorporating the experimentally-measured dielectric function of STO into a phenomenological Eliashberg framework, we find that the dielectric response near 59 meV enhances the pairing interaction and raises the theoretical upper bound of $T_c$ into the experimentally observed range. These results suggest that longitudinal dielectric resonances of polar substrates — a form of resonant anti-shielding (RAS)[40] — can act as electrodynamic amplifiers of pairing interactions and highlight the dielectric environment as an important parameter for understanding, and potentially engineering, superconductivity in interfacial systems.

## Results

### First-principles Eliashberg spectrum and its limitations

Because a fully microscopic treatment of bulk long-range substrate electrodynamics at oxide interfaces remains challenging, we adopt a phenomenological framework designed to isolate the macroscopic longitudinal contribution to the pairing interaction. Our goal is not to construct a complete microscopic theory of FeSe:STO, but rather to identify a generic electrodynamic mechanism that may persist independently of microscopic interface details.

Experimentally, the superconducting properties of FeSe:STO interfaces depend sensitively on growth conditions, including substrate preparation, annealing, and doping. A common feature among high-$T_c$ samples is the presence of a $TiO_x$ layer at the interface.[39,41-43] First-principles DFPT calculations provide a microscopic description of the phonon spectrum and the corresponding Eliashberg function $\alpha^2F(\omega)$ for such interface structures, revealing features in $\alpha^2F(\omega)$ at 59.7, 65.7, and 78.0 meV, all above the ~ 45 meV cutoff of bulk FeSe. This is shown in Fig. 1(a).[39] The latter two peaks dominate the spectrum and arise primarily from out-of-plane oxygen vibrations at the interface. The feature near 59.7 meV contributes only weakly to the total electron–phonon coupling, despite lying close to the FK mode observed experimentally near 59.3 meV.[22]

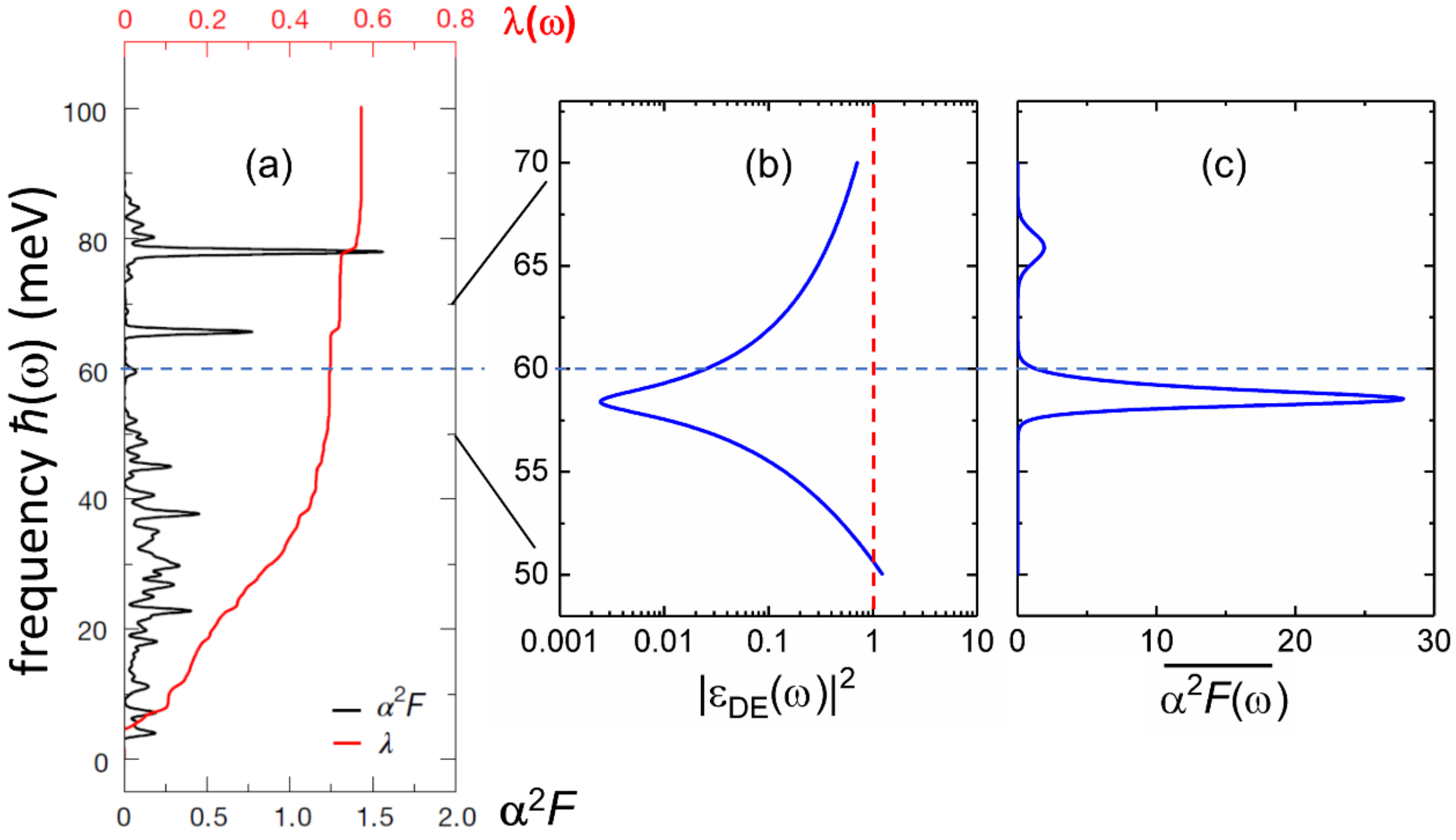

**Fig. 1| Transformation of the Eliashberg function $\alpha^2F(\omega)$ of FeSe monolayer on STO in the presence of RAS.** (a) The Eliashberg function for FeSe:STO taken from Ref. 39. (b) $|\varepsilon(\omega)|^2$ obtained from $\varepsilon(\omega)$ in Fig. 2. [Ref. 55]. (c) The resulting renormalized Eliashberg function $\overline{\alpha^2F(\omega)}$.

These calculations include both interfacial lattice dynamics and charge transfer but are perturbative in nature and therefore do not fully capture the strong resonant amplification of the macroscopic longitudinal field as the bulk dielectric function approaches zero, which we argue underlies the RAS amplification effect. In addition, finite-slab calculations do not fully capture the long-range electrodynamic response associated with dielectric resonances in the bulk substrate. These limitations likely contribute to the relatively small spectral weight associated with the 59 meV feature, in contrast with experimental observations, which identify a pronounced FK mode at nearly the same energy.[22] In practice, the enhancement associated with the dielectric resonance is regularized by finite damping and linewidth broadening in the substrate response.

Despite the interface-specific nature of existing perturbative DFPT Eliashberg calculations, experimental systematics across other polar substrates supporting high temperature superconductivity in monolayer FeSe (Table 1) indicate that the lower FK mode is generically tied to the bulk longitudinal dielectric resonance. This provides a natural explanation for the robustness of superconductivity at these interfaces. These considerations motivate a complementary electrodynamic treatment designed to isolate the macroscopic longitudinal contribution associated with the substrate dielectric response.

**Dielectric enhancement of the Eliashberg interaction**

The coupling between electrons and lattice polarization in polar materials is governed by the longitudinal dielectric response.[44-47] In the long-wavelength limit, the electric field associated with lattice polarization scales as $1/\varepsilon(\omega)$, leading to an electron–phonon matrix element $g(\omega) \propto 1/\varepsilon(\omega)$. Within our phenomenological framework, the pairing kernel entering Migdal-Eliashberg theory[48,49] inherits the dielectric scaling $|g(\omega)|^2 \propto 1/|\varepsilon(\omega)|^2$, resulting in strong enhancement of the phonon-mediated interaction near frequencies where the bulk dielectric function approaches zero. This RAS effect is particularly relevant for the longitudinal optical (LO) resonance of STO, which is closely associated with the lower FK mode at the interface. The present treatment employs the macroscopic dielectric function $\varepsilon(\omega) \equiv \varepsilon(q \to 0, \omega)$. This approximation is appropriate because FK modes correspond to long-wavelength longitudinal polarization at the surface[50] and the associated electron-phonon interaction is dominated by small momentum transfer.[44,3] Our phenomenological approach is described in Methods.

**Upper bound estimate of the superconducting transition temperature**

Within the dielectric framework introduced in Methods, the enhancement of the longitudinal electric field can be incorporated by renormalizing the Eliashberg function in the vicinity of the FK resonance according to

$$\alpha^2F(\omega) \ \rightarrow \ \frac{\alpha^2F(\omega)}{|\varepsilon(\omega)|^2}, \qquad (1)$$

$\varepsilon(\omega)$ is the experimentally-measured macroscopic dielectric function of STO in the long-wavelength limit. The corresponding increment in the $T_c$ upper bound is then estimated using the modified Eliashberg function in a narrow energy window around the FK/LO resonance at 59 meV:

$$\Delta T_c^{\max} = c(\mu^*) \int_{\omega 1}^{\omega 2} \frac{\alpha^2F(\omega)}{|\varepsilon(\omega)|^2}\, d\omega - c(\mu^*) \int_{\omega 1}^{\omega 2} \alpha^2F(\omega)\, d\omega \qquad (2)$$

where the frequency interval extends from $\omega_1 = 50$ meV to $\omega_2 = 70$ meV. Outside this interval the dielectric amplification of $T_c$ is comparatively weak. The resulting upper bound is given by $\overline{T_c^{\max}} = T_c^{\max} + \Delta T_c^{\max}$ where $T_c^{\max}$ is the upper bound without consideration of the coupling to the bulk longitudinal optical phonon modes (see Eq. 3 in Methods). Evaluation of $\Delta T_c^{\max}$ therefore requires two inputs: the Eliashberg function $\alpha^2F(\omega)$ and the dielectric function $\varepsilon(\omega)$. The $T_c$ enhancement is controlled not only by the dielectric resonance itself, but also by its spectral overlap with the Eliashberg function. In practice, the enhancement remains finite due to damping and linewidth broadening in the dielectric response.

When electrons and phonons belong to the same medium, electronic polarization can substantially renormalize the phonon propagator through the phonon self-energy.[46] Here, however, the relevant polar phonon originates predominantly in STO, whereas the additional electronic polarization is confined to the single FeSe layer. We therefore expect the FeSe-induced correction to the STO phonon self-energy to be small compared with the dominant substrate response and neglect this backaction to leading order.

Finally, the efficiency of RAS also depends on the electronic structure of the superconducting layer, since the relevant phonon-mediated scattering processes require suitable electronic states near the Fermi level. Changes in carrier density can modify both the bare electron-phonon coupling and the electronic screening, so the resulting enhancement need not vary monotonically with doping.

**RAS in FeSe:SrTiO$_3$**

We evaluated the effect of the RAS amplification, as described in Methods, with results plotted in Fig. 1. For STO, there are three transverse/longitudinal optical resonances ($j$ = 1,2,3). The dielectric response exhibits a zero crossing near $\omega_{\mathrm{LO}} \approx 59$ meV, producing a strong enhancement in $1/|\varepsilon(\omega)|^2$. Fig. 2(b) shows bulk and surface loss functions at 20 K, $\mathrm{Im}[-1/\varepsilon(\omega)]$ and $\mathrm{Im}[-1/(\varepsilon(\omega)+1)]$, corresponding to longitudinal optical and FK excitations. The lower- energy mode exhibits a strong and sharp resonance near 59 meV, while the higher-energy mode is much broader, indicating a weaker and less sharply peaked dielectric resonance.

Although FK modes formally lie between the transverse and longitudinal optical phonons, their proximity to the LO frequency in strongly polar materials follows directly from the dielectric response. Between the TO and LO frequencies, $\varepsilon(\omega)$ varies rapidly, particularly just above the TO resonance, but its magnitude remains large, $|\varepsilon(\omega)| \gg 1$ over most of this interval as shown in Fig. 2(a). Consequently, the FK condition $\varepsilon(\omega) = -1$ is satisfied only in a narrow frequency window where $\varepsilon(\omega)$ approaches zero, thus constraining the FK mode to lie close to the bulk LO frequency and inherit its longitudinal character. At this frequency, the longitudinal polarization field cannot propagate into the vacuum and is reflected at the interface, enforcing an evanescent field localized near the surface. The resulting surface excitation remains governed primarily by the bulk dielectric response while its electric field extends outside the material. Consequently, this mode carries the strongest longitudinal electric field and provides the dominant channel for electrodynamic coupling to electrons in the FeSe layer.

Only the lower FK mode has substantial overlap with the DFPT Eliashberg function, shown in Fig. 1. The enhancement is controlled by both the dielectric resonance and its spectral overlap with the Eliashberg function. This spectral overlap becomes relevant to the FeSe layer because the electrons couple directly to the evanescent electric field of the FK mode, while the interaction itself remains embedded in the dynamical dielectric environment of the interface. Strictly, the relevant screening is that of the complete FeSe:STO interface; to leading order, however, we approximate it by the bulk STO dielectric function because the dominant polar response resides in STO. The negligible change of the FK modes upon FeSe deposition supports this approximation.

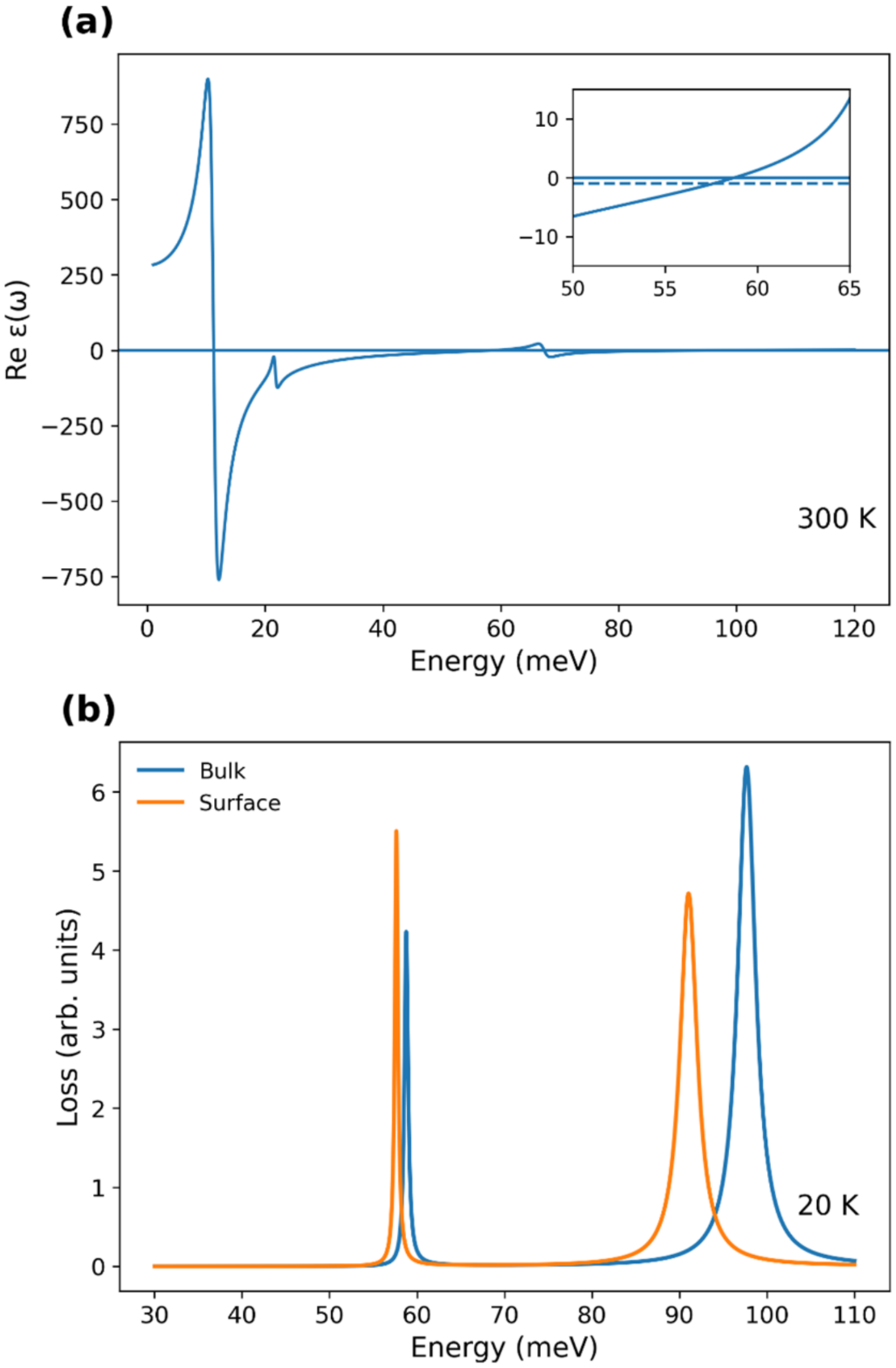

**Fig. 2| Dielectric response and associated loss functions of $SrTiO_3$ (**from Kamarás *et al.* [30]). (a) Real part of the dielectric function, Re $\varepsilon(\omega)$ at 300 K. The inset displays a zoom of the low-temperature (20 K) response in the vicinity of the lower LO resonance. The horizontal lines at $\varepsilon =$

0 and $\varepsilon = -1$ indicate the conditions for bulk longitudinal and surface Fuchs–Kliewer (FK) modes, respectively, illustrating that the FK mode lies close to the bulk LO frequency. (b) Bulk and surface loss functions at 20 K, $\mathrm{Im}[-1/\varepsilon(\omega)]$and $\mathrm{Im}[-1/(\varepsilon(\omega)+1)]$, corresponding to longitudinal optical and FK excitations.

To isolate the effect of this resonance, we fit the calculated region of $\alpha^2F(\omega)$[39] using two Gaussian components, one at 59 meV, and the other at 66 meV to account for another small feature in $\alpha^2F(\omega)$. Next, we divide this by the $|\varepsilon(\omega)|^2$ shown in Fig. 1(b), which produces a pronounced redistribution of spectral weight, transforming the weak 59 meV feature into a sharp, localized peak (Fig. 1(c)). The resulting enhancement reflects the dielectric response in the vicinity of the longitudinal resonance and cannot be fully captured by the finite-slab DFPT treatment. The remainder of the spectrum (*i.e.*, outside the 50-70 meV interval) is left unchanged. The electron–phonon coupling increases from $\lambda \approx 0.57$ to $\tilde{\lambda} \approx 1.1$, while remaining within the conventional moderate-coupling regime.[45] The dielectric-enhanced pairing contribution associated with the 59 meV resonance yields $\Delta T_c^{\max} \approx 77$ K. The baseline DFPT spectrum further increases the upper bound, giving a total $\overline{T_c^{\max}} \approx 96$ K, which is somewhat above the experimentally observed $T_c$ range.

**Alternative implementation using refractive index**

As an independent consistency check of the dielectric enhancement mechanism, we explored an alternative and more direct way to evaluate the dielectric enhancement factor without explicitly reconstructing the full dielectric function. Using $\varepsilon' = n^2 - k^2$ and $\varepsilon'' = 2nk$, and noting that near the relevant longitudinal resonance in STO the extinction coefficient is small ($n^2 \gg k^2$), one obtains $|\varepsilon(\omega)|^2 \approx n^4(\omega)$, where the real part of the refractive index $n(\omega)$ is taken directly from experimental optical data.[55] In the vicinity of the FK/LO resonance, where the dielectric response varies rapidly, $n(\omega)$ can be approximated locally by a linear function, $n(\omega) \approx \alpha(\omega - \omega_k) + \beta$, where $\omega_k$is the resonance frequency. A fit to the experimental data yields $\alpha = 0.90$ and $\beta = 0.11$. This lowest order linear behavior is supported by a diagrammatic analysis.[47] Using this representation, we constructed the renormalized Eliashberg function $\alpha^2F(\omega)$, which again shows a pronounced enhancement near the 59 meV resonance. Applying the Leavens expression (Eq. 2) yields an upper bound of $T_c^{\max} \approx 120$ K, again yielding enhancement on the order of the experimentally observed $T_c$. The persistence of strong $T_c$ enhancement within this simplified implementation indicates that the RAS effect is controlled primarily by the near-resonant dielectric response itself, rather than by the specific parameters entering the dielectric model (Methods, Eq. 6).

**Discussion**

The above analysis shows that macroscopic electrodynamic boundary effects can significantly enhance superconductivity at polar interfaces. Incorporating the measured dielectric response of STO into the pairing kernel strongly enhances the FK/LO channel near 59 meV and redistributes spectral weight within a narrow energy window, an effect we refer to as RAS. The lower LO mode of STO produces a pronounced dielectric zero near 59 meV where the FeSe Eliashberg spectrum already has finite weight, making it particularly effective in enhancing the pairing interaction. By contrast, higher-energy bulk modes contribute less strongly to the Eliashberg spectrum and are therefore less effective in enhancing pairing. In principle, dielectric amplification may also act on these higher-energy modes; however, their weaker spectral overlap with $\alpha^2F(\omega)$ limits their overall contribution within the present framework.

The alignment of the bulk LO modes and surface FK modes is not sufficient to optimize $T_c$; the electronic structure of the superconducting layer must also be considered.[2,3] The optimal carrier concentration for high $T_c$ in FeSe:STO is about $0.1e^-$ per Fe atom. This requires substantial charge transfer across the interface that fills the electron pockets at the M-points.[2] A recent report indicated a doping level of about $0.05e^-$ per Fe for mixed SrO/$TiO_x$-terminated STO surfaces, implying a significantly lower carrier concentration at SrO-terminated interfaces.[56] The reduced carrier density may explain why the superconducting gap of the FeSe monolayer, measured by scanning tunneling microscopy, is strongly suppressed for SrO-terminated STO compared to $TiO_x$-terminated STO, because the bulk LO mode is not expected to shift significantly with surface termination. By contrast, monolayer FeSe on MgO exhibits a modest $T_c$ of 18 K even though the carrier density is of the order of 0.06 to 0.2 $e^-$ per Fe atom, comparable to FeSe:STO.[57] However, MgO only possesses a single polar optical branch: the low-energy FK mode is absent and the remaining high-energy FK phonon at 81 meV is relatively far removed from the bulk LO resonance at 90 meV.[58,59] These comparisons indicate that both the electronic structure of the FeSe layer and the electrodynamic response of the substrate must be considered together in determining the pairing interaction.

Regarding the application of RAS to other systems, the main limitation is the scarcity of sufficiently detailed and reliable DFPT calculations. FeSe/BTO nevertheless provides a useful test case because an *ab initio* Eliashberg function has been reported for this interface, albeit using only a one-unit-cell-thick BTO slab.[6] Using the available optical parameters for tetragonal BTO,[31, 60, 61] we obtain FK (LO) modes near 57 (58) and 82 (89) meV (Table 1), with the corresponding damping parameters taken from Ref. 60. The spectral overlap between the lower LO resonance and the computed $\alpha^2F(\omega)$ spectrum of FeSe/BTO[6] is rather small. By contrast, the upper LO resonance has substantial overlap with $\alpha^2F(\omega)$. The magnitude of this overlap, and hence the resulting RAS enhancement, is sensitive to the precise resonance energy and damping parameters. Applying the same procedure as for FeSe:STO with these parameters yields an upper bound on $T_c$ of about 66 K, comparable with the reported 75 K. Although the limited BTO slab calculation precludes quantitative interpretation, the result provides a useful additional example of RAS.

In our calculations, the estimated $T_c$ is proportional to the integral of $\alpha^2F(\omega)$, and increases from approximately 19 K to values near 100 K, with the corresponding electron-phonon coupling remaining moderate, $\lambda < 1.1$. The resulting enhancement strongly redistributes spectral weight near the longitudinal dielectric resonance while remaining within the conventional moderate-coupling regime. At sufficiently strong enhancement, one may anticipate crossover physics beyond the conventional Migdal-Eliashberg framework, including possible polaronic or bipolaronic effects.[62,63,64] Exploring such regimes lies beyond the scope of the present phenomenological treatment.

The electrodynamic amplification mechanism identified here (RAS) is complementary to forward-scattering scenarios. While forward-focused coupling primarily governs the momentum structure of the interaction and may account for replica bands, dielectric amplification selectively enhances its overall magnitude through macroscopic longitudinal electrodynamics of the substrate. The FeSe:STO system provides a concrete example in which this mechanism can be quantified, while the broader class of FeSe/oxide interfaces suggests that the underlying principle transcends specific lattice structures and chemical compositions. Taken together, these results indicate that dielectric environments can function as electrodynamic pairing amplifiers and can be exploited to enhance superconductivity in engineered heterostructures.

## Methods

### Dielectric enhancement of the Eliashberg interaction

To phenomenologically incorporate the macroscopic electrodynamic contribution associated with the longitudinal dielectric resonance, we adopt a dielectric treatment in which the experimentally-measured bulk dielectric function of STO[30] is included in the pairing kernel. Specifically, the DFPT-derived Eliashberg function $\alpha^2F(\omega)$ is modified within a narrow energy window surrounding the 59 meV FK/LO resonance by a factor $1/|\varepsilon(\omega)|^2$, while the remainder of the spectrum is left unchanged. This selective modification reflects the fact that finite slab DFPT calculations capture local phonon dispersions and electron–phonon coupling, but do not fully reproduce the macroscopic longitudinal electric fields associated with the bulk dielectric zero. Restricting the correction to the vicinity of the dielectric resonance therefore supplements the DFPT spectrum with the dominant electrodynamic contribution while avoiding double counting interactions already incorporated into the first-principles calculation. The analysis is carried out within the Eliashberg framework and therefore applies in the regime of small to moderate electron–phonon coupling.

### Upper bound estimate of the superconducting transition temperature

To estimate the superconducting transition temperature within Eliashberg theory, we employ the formulation proposed by Leavens,[51,45] which provides a rigorous upper bound on $T_c$ within the Eliashberg framework directly in terms of the Eliashberg function $\alpha^2F(\omega)$.[49] This approach has been shown to yield results consistent with full numerical solutions of the Eliashberg equations.[40] The upper bound is given by

$$T_c^{\max} = c(\mu^*) \int_0^\infty \alpha^2\, F(\omega)\, d\omega, \qquad (3)$$

where $c(\mu^*)$ is a weakly decreasing function of the Coulomb pseudopotential $\mu^*$. For typical values $0 \leq \mu^* \leq 0.2$, $c(\mu^*)$ varies from $c(0) \approx 0.23$ to $c(0.2) \approx 0.15$.[45] $T_c^{\max}$ thus depends weakly on this parameter. An equivalent representation is obtained using the cumulative coupling function

$$\lambda(\omega) = 2\int_0^{\omega} \frac{\alpha^2 F(\omega')}{\omega'}\, d\omega', \tag{4}$$

with $\lambda(\infty) \equiv \lambda$,[45] yielding

$$T_c^{\max} = c(\mu^*)\int_0^{\infty} [\lambda(\infty) - \lambda(\omega)]\, d\omega. \tag{5}$$

This form is numerically convenient because $\lambda(\omega)$ is smoother than $\alpha^2 F(\omega)$. The resulting $T_c^{\max}$ should be interpreted as an Eliashberg upper-bound for the pairing interaction defined by the chosen spectral function.

**RAS in FeSe:SrTiO$_3$**

To establish a reference, we first evaluate $T_c^{\max}$ without RAS (Eq. 3). For bulk FeSe, the Eliashberg function $\alpha^2 F(\omega)$ exhibits a broad spectrum spanning approximately 3–45 meV.[52] Digitizing and entering these data[53] in Eqs. (3–5) yields $T_c^{\max} \approx 5.4$ K for $c = 0.23$, in reasonable agreement with the experimental value $T_c \approx 8$ K.[54] The corresponding electron–phonon coupling constant is $\lambda \approx 0.4$. Next, we consider a monolayer of FeSe on STO using the DFPT calculations of $\alpha^2 F(\omega)$ for the periodic slab geometry (Fig. 1).[39] The spectrum contains additional features at 59.7, 65.7, and 78.0 meV, with the latter two dominating. Using Eq. (3–5), we obtain $T_c^{\max} \approx 19$ K. This remains far below the experimentally observed value of 60 – 80 K, indicating that the slab calculation underestimates the contribution of the longitudinal FK/LO resonance.

To evaluate the effect of the RAS amplification, we renormalize the Eliashberg function, using Eq. (1), with the macroscopic dielectric function $\varepsilon(\omega)$ obtained from optical measurements.[30] The dielectric function is modeled using a Lorentz oscillator form

$$\varepsilon(\omega) = \varepsilon_\infty \prod_j^3 \frac{\omega_{LO,j}^2 - \omega^2 + i\gamma_{LO,j}\omega}{\omega_{TO,j}^2 - \omega^2 + i\gamma_{TO,j}\omega} \qquad (6)$$

with parameters fitted to the optical data by Kamaras et al.,[30] and is plotted in Fig. 2(a).
published article.

**Acknowledgements.** K.K. would like to thank Dr. Jan Engelbrecht for stimulating discussions.

**Funding.** H.H.W. was supported by the National Science Foundation Materials Research Science and Engineering Center (MRSEC) program through the UT Knoxville Center for Advanced Materials and Manufacturing under Grant No. DMR-2309083.

| System | Surface | LO1/FK1 Frequency (meV) | Symm. | Character LO1 | LO2/FK2 Frequency (meV) | Symm. | Character LO2 | Comments | Refs. |
|---|---|---|---|---|---|---|---|---|---|
| | | | | | | | | | |
| STO | (001) | 57-59 / 59 | $F_{1u}$ | $TiO_6$ vs. Sr | 98 / 92-97 | $F_{1u}$ | Ti-O stretch | Cubic. $F_{1u}$ spans x, y, z; projection onto [001] | 22-25, 30 |
| BTO | (001) | 58 / 57 | $A_1$+E | $TiO_6$ vs. Ba | 89 / 82 | $A_1$+E | Ti-O stretch | Tetragonal splitting of $F_{1u}$ with $A_1$//[001] | 22, 27, 31 |
| LFO | (001) | 60 / 58 | $B_{1u}$ | Fe-O mixed bending/stretching | 84 / 78 | $B_{1u}$ | Fe-O stretch | Orthorhombic $B_{1u}$ //[001] | 15, 26, 32 |
| ETO* | (001) | 59/ 58 | $F_{1u}$/ $A_{2u}$+$E_u$ | $TiO_6$ vs. Eu | 91/86 | $F_{1u}$/ $A_{2u}$+$E_u$ | Ti-O stretch | Tetragonal splitting of $F_{1u}$ with $A_{2u}$//[001] | 28, 29 |
| Rutile $TiO_2$ | (100) | 45, 55 / 45, 55 | $E_u$ | Ti-O mixed bending/stretching | 99, 103 / 94 | $A_{2u}$, $E_u$ | Ti-O stretch | $E_u$ (103 meV) has component along [100]. $A_{2u}$ $\perp$ [100]. FK modes only measured for (110). | 22, 33, 37 |
| Anatase $TiO_2$ | (001) | 45 / 48 | $E_u$ | Ti-O bending | 94 / 92 | $A_{2u}$ | Ti-O stretch | $A_{2u}$ // [001]; $E_u$ in-plane | 17, 34-36 |

**Table 1| Longitudinal optical (LO) and Fuchs–Kliewer (FK) phonon modes across polar substrates supporting enhanced superconductivity in FeSe monolayers.** LO1 (LO2) denotes the lower (higher) energy bulk longitudinal optical modes, and FK1 (FK2) the corresponding surface excitations. FK modes arise from the reflection of the longitudinal polarization field at the surface and are therefore directly tied to the bulk dielectric response. While all LO modes are polar, the relevant coupling channel is determined by the projection of the longitudinal polarization onto the interface normal. The lower FK/LO mode typically corresponds to the bulk longitudinal dielectric resonance ($\varepsilon(\omega) \to 0$) and carries the strongest macroscopic electric field, providing the dominant channel for electrodynamic coupling to electrons in the FeSe layer. (*) The $EuTiO_3$ LO and FK phonon frequencies were calculated from the optical dielectric-function parameters reported in Refs. [28] with the FK frequencies determined from the surface-mode condition $\epsilon(\omega) + 1 = 0$. The calculated LO frequencies match the reported values in the supplementary information of Ref. 29.

## References

1. Wang, Q.Y. *et al*. Interface-induced high-temperature superconductivity in single unit-cell FeSe films on $SrTiO_3$. *Chin. Phys. Lett.* **29,** 037402 (2012); 10.1088/0256-307X/29/3/037402.

2. He, S. *et al*. Phase diagram and electronic indication of high-temperature superconductivity at 65 K in single-layer FeSe films. *Nat. Mater.* **12**, 605-610 (2013); 10.1038/nmat3648.

3. Lee, J.J. *et al*. Interfacial mode coupling as the origin of the enhancement of $T_c$ in FeSe films on $SrTiO_3$. *Nature* **515**, 245-248 (2014); 10.1038/nature13894.

4. Linscheid, A., Maiti, S., Wang, Y., Johnston, S. & Hirschfeld, P.J. High $T_c$ via spin fluctuations from incipient bands: Application to monolayers and intercalates of FeSe. *Phys. Rev. Lett.* **117**, 077003 (2016); 10.1103/PhysRevLett.117.077003.

5. Li, Z.-X., Wang, F., Yao, H. & Lee, D.-H. What makes the $T_c$ of monolayer FeSe on $SrTiO_3$ so high: A sign-problem-free quantum Monte Carlo study. *Sci. Bull.* **61**, 925-930 (2016); 10.1007/s11434-016-1087-x.

6. Wang, Y., Linscheid, A., Berlijn, T. & Johnston, S. *Ab initio* study of cross-interface electron-phonon couplings in FeSe thin films on $SrTiO_3$ and $BaTiO_3$. *Phys. Rev. B* **93**, 134513 (2016); 10.1103/PhysRevB.93.134513.

7. Zhang, S. *et al.* Enhanced superconducting state in FeSe/$SrTiO_3$ by a dynamic interfacial polaron mechanism. *Phys. Rev. Lett.* **122**, 066802 (2019); 10.1103/PhysRevLett.122.066802.

8. Rademaker, L., Alvarez-Suchini, G., Nakatsukasa, K., Wang, Y. & Johnston, S. Enhanced superconductivity in FeSe/$SrTiO_3$ from the combination of forward scattering phonons and spin fluctuations. *Phys. Rev. B* **103**, 144504 (2021); 10.1103/PhysRevB.103.144504.

9. Pelliciari, J. *et al.* Evolution of spin excitations from bulk to monolayer FeSe, *Nat. Commun.* **12**, 3122; 10.1038/s41467-021-23317-3 (2021).

10. Liu, C. *et al.* High-order replica bands in monolayer FeSe/$SrTiO_3$ revealed by polarization-dependent photoemission spectroscopy. *Nat. Commun.* **12**, 5480; 10.1038/s41467-021-24783-5 (2021).

11. Zhang, C. *et al.* Ubiquitous strong electron–phonon coupling at the interface of FeSe/$SrTiO_3$. *Nat. Commun*. **8**, 14468; 10.1038/ncomms14468 (2017).

12. Rademaker, L., Wang, Y. Berlijn, T. & Johnston, S. Enhanced superconductivity due to forward scattering in FeSe thin films on $SrTiO_3$ substrates. *New J. Phys.* **18**, 022001 (2016); 10.1088/1367-2630/18/2/022001.

13. Zhang, P. *et al.* Observation of high-$T_c$ superconductivity in rectangular FeSe/$SrTiO_3$(110) monolayers. *Phys. Rev.* B **94**, 104510 (2016); 10.1103/PhysRevB.94.104510.

14. Peng, R. Tuning the band structure and superconductivity in single-layer FeSe by interface engineering. *Nat. Commun.* **5,** 5044; 10.1038/ncomms6044 (2014).

15. Song, Y. *et al.* High temperature superconductivity at FeSe/$LaFeO_3$ interface. *Nat. Commun.* **12**, 5926; 10.1038/s41467-021-26201-2 (2021).

16. Rebec, S.N. *et al.* Coexistence of Replica bands and superconductivity in FeSe monolayer films. *Phys. Rev. Lett.* **118**, 067002 (2017); 10.1103/PhysRevLett.118.067002.

17. Ding, H. *et al.* High-temperature superconductivity in single-unit-cell FeSe films on anatase $TiO_2$(001). *Phys. Rev. Lett.* **117**, 067001 (2016); 10.1103/PhysRevLett.117.067001.

18. Liu, C. *et al*. High-temperature superconductivity and its robustness against magnetic polarization in monolayer FeSe on $EuTiO_3$. *npj Quantum Mater.* **6**, 85; 10.1038/s41535-021-00388-5 (2021).

19. Liu, D. *et al*. Electronic origin of high-temperature superconductivity in single layer FeSe superconductor. *Nat. Commun.* **3**, 931; 10.1038/ncomms1946 (2012).

20. Guo, J. *et al.* Superconductivity in the iron selenide $K_xFe_2Se_2$ $(0 \le x \le 1)$. *Phys. Rev. B* **82**, 180520(R) (2010); 10.1103/PhysRevB.82.180520.

21. Lu, X.F. *et al.* Coexistence of superconductivity and antiferromagnetism in $(Li_{0.8}Fe_{0.2})$OHFeSe. *Nat. Mater.* **14**, 325 (2015); 10.1038/nmat4155.

22. Zhang, S. *et al.* Lattice dynamics of ultrathin FeSe films on $SrTiO_3$. *Phys. Rev. B* **97**, 035408 (2018); 10.1103/PhysRevB.97.035408.

23. Zhang, S. *et al.* Role of $SrTiO_3$ phonon penetrating into thin FeSe films in the enhancement of superconductivity. *Phys. Rev. B* **94**, 081116(R) (2016); 10.1103/PhysRevB.94.081116.

24. Cao, Y., Wang, S., Liu, S., Guo, Q. & Guo, J.D. Electronic structures of the $SrTiO_3$(110) surface in different reconstructions. *J. Chem. Phys.* **137**, 044701 (2012); 10.1063/1.4737946.

25. Baden, A.D., Cox, P.A., Egdell, R.G., Orchard A.F. & Willmer, R.J.D. Observation of surface optical phonons on $SrTiO_3$(100). *J. Phys. C* **14**, L1081 (1981); 10.1088/0022-3719/14/34/003.

26. Wührl, F., Rieche, A. Oelschläger, A. Dörr, K. & Widdra, W. Ultrafast electron dynamics upon above band-gap excitation in epitaxial $LaFeO_3$(001) thin films. *New. J. Phys.* **28**, 013501 (2026); 10.1088/1367-2630/ae34db.

27. Premper, J. *et al.* Surface stress and lattice dynamics in oxide ultrathin films. *Phys. Stat. Solidi B* **257**, 1900650 (2020); 10.1002/pssb.201900650.

28. Goian V. *et al*, Antiferrodistortive phase transition in $EuTiO_3$. *Phys. Rev. B* **86**, 054112 (2012); 10.1103/PhysRevB.86.054112.

29. Wadhwa, P., Bosin, A. & Filipetti, A. Giant spin-dependent Seebeck effect from fully spin-polarized carriers in n-doped $EuTiO_3$: a prototype material for spin-caloritronic applications, *J. Mater. Chem. A* **11**, 6842-6853 (2023) (supplementary data); 10.1039/d2ta07197e.

30. Kamarás, K., Reedyk, M., Woollam, J.A. & Tanner, D.B. Infrared reflectivity of $SrTiO_3$. *J. Appl. Phys.* **78**, 1235-1240 (1995); 10.1063/1.360364.

31. Dwij, V. *et al.* Revisiting eigen displacements of tetragonal $BaTiO_3$: Combined first principle and experimental investigation. *Physica B* **624**, 413381 (2022); 10.1016/j.physb.2021.413381.

32. Saha, J. *et al.* Structure, Mossbauer spectroscopy and vibration phonon spectra in valence-bond force-field model approach for distorted perovskites $AFeO_3$ (A = La, Y). *Mat. Chem. Phys*. **240**, 122286 (2020); 10.1016/j.matchemphys.2019.122286.

33. Henderson, M.A. An HREELS and TPD study of water on $TiO_2$(110): The extent of molecular versus dissociative adsorption. *Surf. Sci.* **355**, 151-166 (1996); 10.1016/0039-6028(95)01357-1.

34. Durinck, G., Poelman, H., Clauws, P., Fiermans, L., Vennik, J. & Dalmai, G. Observation of surface phonons on the (001) and (100) surfaces of anatase minerals. *Sol. State Commun.* **80**, 579-581 (1991); 10.1016/0038-1098(91)90155-O.

35. Gonzalez, R.J., Zallen, R. & Berger, H. Infrared reflectivity and lattice fundamentals in anatase $TiO_2$. *Phys. Rev. B* **55**, 7014-7017 (1997); 10.1103/PhysRevB.55.7014.

36. Mikami, M., Nakamura, S., Kitao, O. & Arakawa, H. Lattice dynamics and dielectric properties of $TiO_2$ anatase: a first principles study. *Phys. Rev. B* **66**, 155213 (2002); 10.1103/PhysRevB.66.155213.

37. Schöche, S., Hofmann, T., Korlacki, R., Tiwald, T.E. & Schubert, M. Infrared dielectric anisotropy and phonon modes of rutile $TiO_2$. *J. Appl. Phys.* **113**, 164102 (2013); 10.1063/1.4802715.

38. Rosenstein, B & Shapiro, B. Ya, Phonon mechanism explanation of the superconductivity dichotomy between FeSe and FeS monolayers on $SrTiO_3$ and other substrates. *Phys. Rev. B* **103**, 224517 (2021); 10.1103/PhysRevB.103.224517.

39. Yang, H. *et al*. Phonon modes and electron-phonon coupling at the FeSe/$SrTiO_3$ interface. *Nature* **635**, 332-336 (2024); 10.1038/s41586-024-08118-0.

40. Kempa, K., Protik, N.H., Dodge, T., Draxl, C. & Naughton, M.J. Enhancement of superconductivity via resonant anti-shielding with topological plasmon-polarons. *Phys. Rev. B* **107**, 184518 (2023); 10.1103/PhysRevB.107.184518.

41. Li, F. *et al*. Atomically resolved FeSe/ $SrTiO_3$(001) interface structure by scanning transmission electron microscopy. *2D Mat.* **3**, 024002; 10.1088/2053-1583/3/2/024002 (2016).

42. Sims, H. *et al*. Intrinsic interfacial van der Waals monolayers and their effect on the high-temperature superconductor FeSe/$SrTiO_3$. *Phys. Rev. B* **100**, 144103 (2019); 10.1103/PhysRevB.100.144103.

43. Peng, R. *et al*. Picoscale structural insight into superconductivity of monolayer FeSe/$SrTiO_3$. *Sci. Adv.* **6**, eaay4517; 10.1126/sciadv.aay4517 (2020).

44. Fröhlich, H. Theory of the superconducting state. I. The ground state at the absolute zero of temperature. *Phys. Rev.* **79**, 845-855 (1950); 10.1103/PhysRev.79.845.

45. Marsiglio, E.F. & Carbotte, J.P. Electron-phonon superconductivity, in Superconductivity (ed. Bennemann, K.H. Ketterson J.B.) 73-162 (Springer, 2008).

46. Mahan, G.D. Many-particle physics (Plenum, 1990).

47. Mattuck, R.D. A guide to Feynman diagrams in the many-body problem (Dover, 1976).

48. Migdal, A.B. Interaction between electrons and lattice vibrations in a normal metal. *JETP* **7**, 996-1001 (1958).

49. Eliashberg, G.M. Interactions between electrons and lattice vibrations in a superconductor. *JETP* **11**, 696-702 (1960).

50. Fuchs H. & Kliewer, K.L. Optical modes of vibration in an ionic crystal slab. *Phys. Rev.* **140**, A2076 (1965); 10.1103/PhysRev.140.A2076.

51. Leavens, C.R. A least upper bound on the superconducting transition temperature. *Solid State Commun.* **17**, 1499-1504 (1975); 10.1016/0038-1098(75)90982-5.

52. Wang, W., Sun, J.F., Li, S.W. & H.Y. Lu. Electronic structure and phonon spectrum of binary iron-based superconductor FeSe in both nonmagnetic and striped antiferromagnetic phases. *Physica* C **472**, 29-33 (2012); 10.1016/j.physc.2011.10.004.

53. Webplot Digitizer. https://apps.automeris.io/wpd4.

54. Hsu, F.-C., Luo, J.-Y., Yeh, K.-W. & Wu, M.-K. Superconductivity in the PbO-type structure α-FeSe. *Proc. Natl. Acad. Sci.* **105**, 14262-14264 (2008); 10.1073/pnas.0807325105.

55. Maletic, S., Maletic, D., Petronijevic, I., Dojcilovic J. & Popovic, D.M. Dielectric and infrared properties of $SrTiO_3$ single crystal doped by 3d (V, Mn, Fe, Ni) and 4f (Nd, Sm, Er) ions. *Chin. Phys. B* **23**, 026102 (2014); 10.1088/1674-1056/23/2/026102.

56. Zou, Q. *et al*. Unraveling enhanced superconductivity in single-layer FeSe through substrate surface terminations. *Nano Lett.* **25**, 16572-16579 (2025); 10.1021/acs.nanolett.5c01298.

57. Zhou, G. *et al*. Interface enhanced superconductivity in monolayer FeSe films on MgO(001): Charge transfer with atomic substitution. *Sci. Bull.* **63**, 747-752 (2018); 10.1016/j.scib.2018.05.016.

58. Thiry, P.A., Liehr, M., Pireaux J.J. & Caudano, R. Infrared optical constants of insulators determined by high-resolution electron-energy-loss spectroscopy. *Phys. Rev. B* **29**, 4824-4826 (1984); 10.1103/PhysRevB.29.4824.

59. Sangster, M.J.L., Peckham G. & Saunderson, D.H. Lattice dynamics of MgO. *J. Phys. C* **3**, 1026-1036 (1970); 10.1088/0022-3719/3/5/017.

60. Servoin, J.L., Gervais, F., Quittet, A.M. & Luspin, Y. Infrared and Raman responses in ferroelectric perovskite crystals: Apparent inconsistencies, *Phys. Rev. B* **21**, 2038-2041 (1980); 10.1103/PhysRevB.21.2038.

61. Chavez, A., Katiyar, R.S. & Porto, S.P.S. Coupled modes with $A_1$ symmetry in tetragonal $BaTiO_3$. *Phys. Rev. B* **10**, 3522-3533 (1974); 10.1103/PhysRevB.10.3522.

62. Leggett, A.J., in Modern Trends in the Theory of Condensed Matter, (ed. Pekalski A. & Przystawa, R. Springer-Verlag, 1980).

63. Zhang, J., Sous, J. Reichman, D., Berciu, M., Millis, A.J. Prokof'ev, N.V. & Svistunov, B.V. Bipolaronic high-temperature superconductivity. *Phys. Rev. X* **13**, 011010; 10.1103/PhysRevX.13.011010 (2023).

64. Qin, W. & Yang, Y-F. Intrinsic constraint on $T_c$ for unconventional superconductivity. npj Quantum Mater. **10**, 13; 10.1038/s41535-025-00733-y (2025).